\documentclass[12pt]{iopart}

\usepackage{iopams}

\usepackage[english]{babel}

\newcommand{\dd}{\mbox{d}}

\begin{document}

\title[Fractional L\'evy motion through path integrals]{Fractional
L\'evy motion through path integrals}

\author{I. Calvo$^1$\footnote{Corresponding author.}, R. S\'anchez$^2$,
and B. A. Carreras$^3$}

\address{$^1$ Laboratorio Nacional de Fusi\'on, Asociaci\'on
EURATOM-CIEMAT, 28040 Madrid, Spain}

\address{$^2$ Fusion Energy Division, Oak Ridge National Laboratory,
Oak Ridge, TN 37831, U.S.A.}

\address{$^3$ BACV Solutions Inc., Oak Ridge, TN 37830, U.S.A.}

\eads{\mailto{ivan.calvo@ciemat.es}, \mailto{sanchezferlr@ornl.gov},
\mailto{bacv@comcast.net}}

\begin{abstract}
Fractional L\'evy motion (fLm) is the natural generalization of
fractional Brownian motion in the context of self-similar stochastic
processes and stable probability distributions. In this paper we give
an explicit derivation of the propagator of fLm by using path integral
methods. The propagators of Brownian motion and fractional Brownian
motion are recovered as particular cases. The fractional diffusion
equation corresponding to fLm is also obtained.
\end{abstract}

\pacs{02.50.Ey, 05.40.Jc, 05.40.Fb}

\section{Introduction}

It is widely known that if $\xi_2(t)$ is a Gaussian, uncorrelated
noise (i.e. white noise) the Langevin (stochastic) equation
\begin{equation}\label{eq:LangBM}
x(t) = x_0 +\int_0^t\xi_2(t')\dd t',
\end{equation}
describes ordinary Brownian motion. One of the properties of Brownian
motion is that the average squared displacement grows linearly with
time, $\langle(x(t)-x_0)^2\rangle\propto t$. However, many transport
processes in physical, biological and social systems exhibit anomalous
diffusion~\cite{Metzler00,Zaslavsky02}. That is,
$\langle(x(t)-x_0)^2\rangle\propto t^{2H}$, with $H\neq 1/2$, where
$H$ is called the Hurst exponent~\cite{Hurst}. In the past, several
authors have attempted to generalize Eq.~(\ref{eq:LangBM}) in order to
accommodate these anomalous processes. The anomalous behaviour may be
associated to the existence of spatiotemporal correlations that
produce correlated increments $\dd x(t):= x(t+\dd t) - x(t)$. For that
reason, the first proposed generalization had the form
\begin{equation}\label{eq:LangfBM}
x(t) = x_0 +  \frac{1}{\Gamma(H+1/2)}\int_0^t (t-t')^{H -
1/2}\xi_2(t')\dd t',
\end{equation}
known as fractional Brownian motion (fBm)~\cite{Mandelbrot68}, which
has been extensively studied and applied
\cite{Molz97,book:Elliott,Sanchez06,Cherayil}. In
Eq.~(\ref{eq:LangfBM}) the noise $\xi_2(t)$ remains Gaussian and
uncorrelated but, thanks to the convolution with the power-law kernel,
the increments $\dd x(t)$ become correlated in a way that yields the
desired scaling of the mean squared displacement as well as other
self-similar properties. In particular, the average motion remains
invariant under the transformation $(x,t)\mapsto (\lambda^H x, \lambda
t)$, generalizing the self-similarity of the original Brownian motion,
which is obviously recovered when $H=1/2$ is set in
Eq.~(\ref{eq:LangfBM}). $H$ is the self-similarity exponent of the
process. When rewritten in terms of the Riemann-Liouville fractional
integral operators (see \ref{sec:FDOrealine}) fBm reads
\begin{equation}\label{eq:LangfBM2}
x(t) = x_0 +  {}_0D_t^{-(H + 1/2)}\xi_2.
\end{equation}
The corresponding propagator and diffusion equation of fBm have been
derived in a number of ways in the literature (see, for example,
\cite{Sebastian95} and \cite{Wang90}).

We will devote this paper to the L\'evy generalization of fractional
Brownian motion, known as fractional L\'evy motion
(fLm)~\cite{Laskin02,Huillet,Painter94,Painter96}. The Langevin
equation defining the process is
\begin{equation}\label{eq:LangevinEq}
x(t) = x_0 + {}_0D_t^{-H + 1/\alpha - 1}\xi_\alpha,
\end{equation}
where $\xi_\alpha(t)$ is time-uncorrelated and distributed, for each
$t$, according to a symmetric L\'evy distribution~\cite{Taqqu}. We
recall here that symmetric L\'evy distributions are the symmetric
solutions of the Generalized Central Limit Theorem and are
parametrized by the stability index, $\alpha\in(0,2]$, and the scale
factor, $\sigma>0$. The characteristic function (i.e. the Fourier
transform) of a symmetric L\'evy distribution $L_{\alpha,\sigma}(u)$
is
\begin{equation}\label{eq:Levypdf}
{\cal F}[L_{\alpha,\sigma}](k) = \exp(-\sigma^\alpha|k|^\alpha).
\end{equation}
In particular, for $\alpha\in(0,2)$, the L\'evy distributions have
algebraic tails,
\begin{equation}
L_{\alpha,\sigma}(u) \sim \frac{C_\alpha}{|u|^{\alpha+1}}, \quad |u|\to\infty.
\end{equation}
A L\'evy distribution with $\alpha=2$ is a Gaussian,
\begin{equation}\label{eq:Gaussianpdf}
L_{2,\sigma}(u)
=\frac{1}{2\sigma\sqrt{\pi}}\exp\left(-\frac{u^2}{4\sigma^2}\right),
\end{equation}
and $\sigma$ is related to the second moment, $\langle u^2\rangle =
2\sigma^2$.

The need for fLm originates in the observation of power-law (L\'evy)
statistics for the displacements $\dd x(t)$ in many physical systems
of interest~\cite{Metzler00,Zaslavsky02}, which is in contrast to the
Gaussian character of fBm. In Eq.~(\ref{eq:LangevinEq}), $H$ is still
the self-similarity exponent of the process. It can be
shown~\cite{Taqqu} that finiteness requirements for certain moments of
$x$ restrict the admissible values of $H$ to
\begin{eqnarray}
&&H\in
\left \{
\begin{array}{cc}
(0,\frac{1}{\alpha}]&0<\alpha \le 1\\[4pt]
(0,1]&1 < \alpha \le 2.
\end{array}
\right .
\end{eqnarray}
Under these circumstances the $s$-th moment of $x$, with $0<s<\alpha$,
behaves as $\langle |x|^s\rangle \propto t^{s H}$.

Note that fBm is recovered from Eq.~(\ref{eq:LangevinEq}) if
$\alpha=2$.  Also, the increments $\dd x(t)$ of the process are
uncorrelated for $H=1/\alpha$. In this case, ordinary L\'evy motion is
obtained, of which Brownian motion is a particular case
($H=1/\alpha=1/2$). In the present work we will compute in detail the
propagator of fLm through path-integral
techniques~\cite{book:FeynmanHibbs}, now familiar in both quantum
field theory and statistical physics. Although the form of this
propagator has been previously derived in a more abstract way by using
self-similarity and stability arguments~\cite{Laskin02}, the
path-integral calculation offers a new insight which might help extend
the range of applications of fLm and even tackle more complicated
situations. Two of us recently computed the propagator of fBm by path
integral methods~\cite{CalvoSanchez08} taking advantage of the fact
that the measure can be written as the exponential of an action
quadratic in the fields. As we will see the fLm path-integral measure
is not Gaussian in the fields (except for $\alpha=2$) and the
computation becomes quite more involved.

The rest of the paper is organized as follows. In Section
\ref{sec:ConstMeasure} we construct the appropriate probability
measure on the space of realizations of the noise for fLm. The
propagator is defined as a particular expectation value. Section
\ref{sec:PropagatorPathInt} gives a detailed calculation of the
propagator of fLm in the path integral formalism. In Section
\ref{sec:FracDiffEqs} the fractional diffusion equation satisfied by
the propagator of fLm is worked out. Section \ref{sec:Conclusions}
contains the conclusions. \ref{sec:FDOrealine} collects some basic
definitions on fractional integrals and derivatives.

\section{Construction of the path integral measure}
\label{sec:ConstMeasure}

Assume that the motion of a particle is defined by a stochastic
differential equation. The propagator $G(x_T,T|x_0,0)$ is, by
definition, the probability to find the particle at $x=x_T$ at time
$t=T$ if initially, $t=0$, it was located at $x=x_0$. As mentioned above,
the main objective of this paper is to compute the propagator
associated to Eq.~(\ref{eq:LangevinEq}) (we drop the subscript $\alpha$ of
$\xi$ from now on) by means of path integrals. Consider trajectories
$x(t):[0,T]\to{\mathbb R}$ with boundary conditions $x(0)=x_0$ and
$x(T)=x_T$. From Eq.~(\ref{eq:LangevinEq}) we immediately deduce that the
boundary conditions of $x(t)$ are translated into the following
constraint on $\xi(t)$:
\begin{equation}\label{eq:ConstNoise}
{}_0D_T^{-H + 1/\alpha - 1}\xi = x_T-x_0.
\end{equation}

The essential object in the path integral formalism is the probability
measure ${\cal P}(\xi(t)){\cal D}\xi(t)$ on the space of realizations
of the noise, i.e. on the space of maps $\xi(t):[0,T]\to{\mathbb
R}$. Once it is constructed, the propagator is defined as the
following expectation value:
\begin{equation}\label{eq:defpropagator}
G(x_T,T|x_0,0) =
\int
\delta\left({}_0D_T^{-H + 1/\alpha - 1}\xi - (x_T-x_0)\right)
{\cal P}(\xi(t)){\cal D}\xi(t).
\end{equation}

In order to construct the measure associated to the Langevin equation
(\ref{eq:LangevinEq}) we will discretize the time in $N+1$ points $t_n
:= n\epsilon$, $n=0,1,\dots,N$, with $\epsilon:=T/N$. The continuum
limit, $N\to\infty$, will be taken eventually. Each path is
discretized according to $x_n := x(t_n)$. The appropriate
discretization of the noise is made by taking
$\xi(t_n)=\epsilon^{-1+1/\alpha}\xi_n$, where each $\xi_n$ is an
independent random variable distributed according to a symmetric
L\'evy distribution of index $\alpha$. The factor
$\epsilon^{-1+1/\alpha}$ ensures the correct time-dependence of the
finite moments of $x$, $\langle |x|^s\rangle \propto t^{s
H},~0<s<\alpha$. Therefore, the probability measure is naturally
defined as
\begin{equation}
{\cal P}(\xi(t)){\cal D}\xi(t) = \prod_{n=1}^{N}
L_{\alpha,\sigma}({\xi}_n)\dd{\xi}_n.
\end{equation}

Using the definition of the fractional integral, Eq.~(\ref{RL0}), the
constraint (\ref{eq:ConstNoise}) can be written as
\begin{equation}\label{eq:BCatT}
\frac{1}{\Gamma(H-1/\alpha+1)}\int_0^T(T-\tau)^{H-1/\alpha}{\xi}(\tau)\dd\tau
= x_T-x_0.
\end{equation}
Discretizing as prescribed above:
\begin{eqnarray}
\fl\int_0^T(T-\tau)^{H-1/\alpha}{\xi}(\tau)\dd\tau &=& \sum_{n=1}^N
{\xi}(n\epsilon)\int_{(n-1)\epsilon}^{n\epsilon}
(T-\tau)^{H-1/\alpha}\dd\tau\\[5pt]
&=& \frac{\epsilon^{H}}{H-1/\alpha+1}\sum_{n=1}^N {\xi}_n\left[
(N-n+1)^{H-1/\alpha+1} - (N-n)^{H-1/\alpha+1}
\right],\nonumber
\end{eqnarray}
and we can easily solve for ${\xi}_N$ in terms of ${\xi}_n$,
$n=1,\dots,N-1$:
\begin{equation}
{\xi}_N = A - \sum_{n=1}^{N-1}B_n {\xi}_n,
\end{equation}
with
\begin{eqnarray}
A&:=&
\frac{\Gamma(H-1/\alpha+2)}{\epsilon^{H}}(x_T-x_0),\nonumber\\[5pt]
B_n &:=& (N-n+1)^{H-1/\alpha+1}-(N-n)^{H-1/\alpha+1}, \ \
n=1,\dots, N-1.
\end{eqnarray}

Now, we are ready to write an explicit expression for the expectation
value defining the propagator (\ref{eq:defpropagator}). Namely,
\begin{equation}\label{eq:pathintzvar2withNvar}
\fl G(x_T,T|x_0,0) = \lim_{N\to\infty} f(T,N)\int\delta\left(\xi_N-A +
\sum_{n=1}^{N-1}B_n {\xi}_n\right)\prod_{n=1}^{N}
L_{\alpha,\sigma}({\xi}_n)\dd{\xi}_n,
\end{equation}
where $f(T,N)$ is a normalization factor which will be determined at
the end of the calculation. Equivalently, integrating over $\xi_N$ in
Eq.~(\ref{eq:pathintzvar2withNvar}):
\begin{equation}\label{eq:pathintzvar2}
\fl G(x_T,T|x_0,0) = \lim_{N\to\infty} f(T,N)\int
L_{\alpha,\sigma}\left(A - \sum_{n=1}^{N-1}B_n {\xi}_n\right)
\prod_{n=1}^{N-1}L_{\alpha,\sigma}({\xi}_n)\dd {\xi}_n.
\end{equation}

It is instructive to show that the path integral of ordinary Brownian
motion, usually introduced in a different fashion, coincides with
Eq.~(\ref{eq:pathintzvar2withNvar}) when $H=1/\alpha=1/2$. The Langevin
equation for Brownian motion is (recall Eq.~(\ref{eq:LangBM}))
\begin{equation}
{\dot x}(t)=\xi_2(t).
\end{equation}

The propagator is customarily introduced as
\begin{equation}
\fl G(x_T,T|x_0,0) = \int
\delta(x(0)-x_0)\delta(x(T)-x_T)
\exp\left(-\frac{1}{4\sigma^2}\int_0^T\dot x(t)^2\dd t\right)
{\cal D}x(t),
\end{equation}
where the paths are weighted by the classical action of the free
particle. Now, one can choose the velocity, $v(t)=\dot x(t)$, as the
integration variable. The transformation is linear and the Jacobian
does not depend on the fields. The boundary conditions are simply
translated into
\begin{equation}
\int_0^T v(t)\dd t = x_T - x_0.
\end{equation}
Therefore, we can write
\begin{equation}
\fl G(x_T,T|x_0,0) = \int\delta\left(\int_0^T v(t)\dd t - (x_T -
x_0)\right)\exp\left(-\frac{1}{4\sigma^2}\int_0^Tv(t)^2\dd t\right) {\cal
D}v(t).
\end{equation}
If we discretize the paths as above we get
\begin{equation}
\fl G(x_T,T|x_0,0) =
\lim_{N\to\infty}f(T,N)\int\delta\left(\epsilon\sum_{n=1}^N v_n -
(x_T -
x_0)\right) \prod_{n=1}^N\exp\left(-\frac{\epsilon}{4\sigma^2}
v_n^2\right)\dd v_n,
\end{equation}
where $f(T,N)$ is a normalization factor. Finally, with a last change
of variables, $\xi_n:=\epsilon^{1/2}v_n$ (redefine $f(T,N)$ as
needed):
\begin{equation}
\fl G(x_T,T|x_0,0) = \lim_{N\to\infty}f(T,N)\int\delta\left(\sum_{n=1}^N
\xi_n - \frac{x_T - x_0}{\epsilon^{1/2}}\right)\prod_{n=1}^N
\exp\left(-\frac{1}{4\sigma^2}\xi_n^2\right) \dd \xi_n,
\end{equation}
which is exactly Eq.~(\ref{eq:pathintzvar2withNvar}) for
$H=1/\alpha=1/2$ (recall Eq.~(\ref{eq:Gaussianpdf})).

\section{Path integral computation of the propagator of fractional L\'evy motion}
\label{sec:PropagatorPathInt}

The computation of (\ref{eq:pathintzvar2}) is performed by repeated
use of the identity
\begin{equation}\label{eq:PropConv}
\fl \int_{-\infty}^\infty L_{\alpha,\sigma}(x)L_{\alpha,\sigma}(y-\lambda
x)\dd x = (1+\lambda^\alpha)^{-1/\alpha
}L_{\alpha,\sigma}\left(\frac{y}{(1+\lambda^\alpha)^{1/\alpha}}\right),
\ \forall y\in\mathbb R,
\end{equation}
where $L_{\alpha,\sigma}(x)$ is a symmetric L\'evy distribution with
index $\alpha$ and scale factor $\sigma$, and $\lambda$ is a positive
real number. The proof is straightforward. Define $\bar
L_{\alpha,\sigma}(x):=L_{\alpha,\sigma}(\lambda x)$. Then,
\begin{equation}\label{eq:PropConv2}
\int_{-\infty}^\infty L_{\alpha,\sigma}(x)L_{\alpha,\sigma}(y-\lambda
x)\dd x =  {\cal F}^{-1}[{\cal
F}[L_{\alpha,\sigma}]{\cal F}[\bar L_{\alpha,\sigma}]](\lambda^{-1}y).
\end{equation}
Using that ${\cal F}[\bar L_{\alpha,\sigma}](k)=|\lambda|^{-1}{\cal
F}[L_{\alpha,\sigma}](k/\lambda)$ and ${\cal
F}[L_{\alpha,\sigma}](k)=\exp(-\sigma^\alpha|k|^\alpha)$ we get:
\begin{equation}
{\cal F}[L_{\alpha,\sigma}](k){\cal F}[\bar L_{\alpha,\sigma}](k)=
|\lambda|^{-1}\exp(-\sigma^\alpha|(1 +
|\lambda|^{-\alpha})^{1/\alpha}k|^\alpha).
\end{equation}
And Eq.~(\ref{eq:PropConv}) follows easily.

\vskip 0.5cm

Let us go back to Eq.~(\ref{eq:pathintzvar2}). Using Eq.~(\ref{eq:PropConv})
we integrate out ${\xi}_1$:
\begin{equation}\label{eq:pathintzvar3}
\fl G(x_T,T|x_0,0) = \lim_{N\to\infty} f(T,N)\int
L_{\alpha,\sigma}\left(\frac{A - \sum_{n=2}^{N-1}B_n
{\xi}_n}{(1+B_1^\alpha)^{1/\alpha}}\right)\prod_{n=2}^{N-1}
L_{\alpha,\sigma}({\xi}_n)\dd {\xi}_n.
\end{equation}
Integration of ${\xi}_2$ yields
\begin{eqnarray}\label{eq:pathintzvar4}
\fl G(x_T,T|x_0,0) &=& \lim_{N\to\infty} f(T,N)\int
L_{\alpha,\sigma}\left(\frac{A - \sum_{n=3}^{N-1}B_n
{\xi}_n}{(1+B_1^\alpha)^{1/\alpha}\left(1+\frac{B_2^\alpha}{1+
B_1^\alpha}\right)^{1/\alpha}}\right)\prod_{n=3}^{N-1}
L_{\alpha,\sigma}({\xi}_n)\dd {\xi}_n\nonumber\\[5pt] &=&
\lim_{N\to\infty} f(T,N)\int L_{\alpha,\sigma}\left(\frac{A -
\sum_{n=3}^{N-1}B_n {\xi}_n}{(1+B_1^\alpha
+B_2^\alpha)^{1/\alpha}}\right)\prod_{n=3}^{N-1}
L_{\alpha,\sigma}({\xi}_n)\dd {\xi}_n.
\end{eqnarray}
And after $N-1$ integrations:
\begin{equation}\label{eq:LevyTimeDep}
G(x_T,T|x_0,0) = \lim_{N\to\infty} f(T,N)L_{\alpha,\sigma}\left(
 \frac{A}{(1+\sum_{n=1}^{N-1}B_n^\alpha)^{1/\alpha}}\right).
\end{equation}

It remains to compute
\begin{eqnarray}\label{eq:LevyTimeDep2}
&& \lim_{N\to\infty}\frac{A}{(1+\sum_{n=1}^{N-1}B_n^\alpha)^{1/\alpha}} =
\lim_{N\to\infty}\Gamma(H-1/\alpha+2)\frac{x_T-x_0}{T^H}\times\nonumber\\[5pt]
&& N^H\left[1 + \sum_{n=1}^{N-1}\left((N-n+1)^{H-1/\alpha+1}-
(N-n)^{H-1/\alpha+1}\right)^\alpha\right]^{-1/\alpha}.
\end{eqnarray}

In the sequel, $g(N)\sim h(N)$ will mean that $g(N)/h(N)\to 1$ when
$N\to\infty$. First observe that
\begin{eqnarray}\label{eq:Limit1}
\fl&& \sum_{n=1}^{N-1}\left((N-n+1)^{H-1/\alpha+1}-
(N-n)^{H-1/\alpha+1}\right)^\alpha \sim\cr
&&(H-1/\alpha+1)^\alpha N^{\alpha
H-1}\sum_{n=1}^{N-1}\left(1-\frac{n}{N}\right)^{\alpha H-1},
\end{eqnarray}
where we have used 
\begin{equation}
\fl \left(1-\frac{n}{N}+\frac{1}{N}\right)^{H-1/\alpha+1} \sim
\left(1-\frac{n}{N}\right)^{H-1/\alpha+1} +
\frac{H-1/\alpha+1}{N}\left(1-\frac{n}{N}\right)^{H-1/\alpha}.
\end{equation}
Now, we note that
\begin{equation}\label{eq:EquivSumInt}
\fl \sum_{n=1}^{N-1}\left(1-\frac{n}{N}\right)^{\alpha H-1}\frac{1}{N}\sim
\int_{1/N}^{1-1/N}(1-u)^{\alpha H -1}\dd u = \frac{1}{\alpha
H}\left[\left(1-\frac{1}{N}\right)^{\alpha H}-\frac{1}{N^{\alpha H}}\right],
\end{equation}
and we have almost reached our goal. Combining Eqs.~(\ref{eq:Limit1})
and (\ref{eq:EquivSumInt}) we get
\begin{equation}
\fl \sum_{n=1}^{N-1}\left((N-n+1)^{H-1/\alpha+1}-
(N-n)^{H-1/\alpha+1}\right)^\alpha \sim
\frac{(H-1/\alpha+1)^\alpha}{\alpha H}N^{\alpha H}.
\end{equation}
Inserting this in Eq.~(\ref{eq:LevyTimeDep2}):
\begin{equation}
 \lim_{N\to\infty} \frac{A}{(1+\sum_{n=1}^{N-1}B_n^\alpha)^{1/\alpha}}=
(\alpha
H)^{1/\alpha}\Gamma(H-1/\alpha+1)\frac{x_T-x_0}{T^H}.
\end{equation}

Hence,
\begin{equation}\label{eq:FinalResProp}
G(x_T,T|x_0,0) =
f(T)L_{\alpha,\sigma}\left((\alpha
H)^{1/\alpha}\Gamma(H-1/\alpha+1)\frac{x_T-x_0}{T^H}\right),
\end{equation}
and $f(T)$ can be determined by normalization, $\int_{-\infty}^\infty
G(x_T,T|x_0,0)\dd x_T=1$:
\begin{equation}
f(T)=\frac{(\alpha H)^{1/\alpha}\Gamma(H-1/\alpha+1)}{T^H},
\end{equation}
so that the final expression of the propagator is
\begin{equation}\label{eq:FinalResProp2}
\fl G(x_T,T|x_0,0) = \frac{(\alpha
H)^{1/\alpha}\Gamma(H-1/\alpha+1)}{T^H}L_{\alpha,\sigma}\left((\alpha
H)^{1/\alpha}\Gamma(H-1/\alpha+1)\frac{x_T-x_0}{T^H}\right).
\end{equation}

Summarizing, we have obtained that the propagator of fLm is a L\'evy
distribution depending on the combination $x/t^H$, so that the average
motion is self-similar with exponent $H$.

\section{Fractional diffusion equation}
\label{sec:FracDiffEqs}

For the sake of completeness we derive in this section the fractional
diffusion equation which governs the time evolution of the propagator
of fLm. Denote by $\hat G(k,t)$ the Fourier transform of
$G(x,t|x_0,0)$ with respect to $x$. Using
Eq.~(\ref{eq:FinalResProp2}), the form of the characteristic function
of a L\'evy distribution, and the properties of the Fourier transform
under rescaling:
\begin{equation}
\hat G(k,t) = \exp\left(- \frac{\sigma^\alpha t^{\alpha H}}{(\alpha H)
\Gamma^{{}^\alpha}(H-1/\alpha+1)}|k|^\alpha\right).
\end{equation}

Differentiating with respect to $t$:
\begin{equation}
\frac{\partial}{\partial t} \hat G(k,t) = - \frac{\sigma^\alpha t^{\alpha
H-1}}{\Gamma^{{}^\alpha}(H-1/\alpha+1)}|k|^\alpha \hat G(k,t).
\end{equation}
Fourier inverting, recalling the definition Eq.~(\ref{eq:Riesz}) and the
identity
\begin{equation}
{\cal F}\left[\frac{\partial^\alpha f}{\partial|x|^\alpha}\right](k) =
-|k|^\alpha f(k),
\end{equation}
we find:
\begin{equation}\label{eq:kineqfLm}
\frac{\partial}{\partial t} G(x,t) = \frac{\sigma^\alpha t^{\alpha H-1}}{
\Gamma^{{}^\alpha}(H-1/\alpha+1)} \frac{\partial^\alpha
}{\partial|x|^\alpha}G(x,t).
\end{equation}
Therefore, the propagator of fLm satisfies a space-fractional
diffusion equation with time-dependent
diffusivity. Eq.~(\ref{eq:kineqfLm}) was recently derived by different
methods in \cite{Watkins08}.

The equation for the propagator of fBm (appeared in \cite{Wang90}) is
obtained from Eq.~(\ref{eq:kineqfLm}) in the particular case
$\alpha=2$:
\begin{equation}
\frac{\partial}{\partial t} G(x,t) = \frac{\sigma^2 t^{2H-1}}{
\Gamma^{{}^2}(H+1/2)} \frac{\partial^2 }{\partial x^2}G(x,t),
\end{equation}
which is a diffusion equation with time-dependent diffusivity.

Finally, if $H=1/\alpha=1/2$ we retrieve the standard diffusion
equation associated to ordinary Brownian motion:
\begin{equation}
\frac{\partial}{\partial t} G(x,t) =  \sigma^2
\frac{\partial^2 }{\partial x^2}G(x,t).
\end{equation}

\section{Conclusions}\label{sec:Conclusions}

The Langevin equation defining fLm consists of two main ingredients: a
time-uncorrelated stochastic noise distributed according to a L\'evy
distribution and a fractional integral operator which generates the
time correlations. In this paper we have derived the propagator of fLm
(which was deduced in \cite{Laskin02} by using self-similarity
arguments) through path integral techniques. That is, we have
explicitly constructed a probability measure on the set of
realizations of the noise and precisely defined the propagator as an
average over this measure space. The computation of the propagator has
been performed by discretizing the paths and carefully taking the
continuum limit at the end. The fractional diffusion equation
associated to fLm has also been derived. We hope that the heuristic
power of the path integral formalism will provide new insight on the
calculation and help address more complicated cases.

\vskip 0.2cm

{\bf Acknowledgements:} I. C. acknowledges the hospitality of Oak
Ridge National Laboratory, where this work was carried out. Part of
this research was sponsored by the Laboratory Research and Development
Program of Oak Ridge National Laboratory, managed by UT-Battelle, LLC,
for the US Department of Energy under contract number
DE-AC05-00OR22725.  B. A. C. acknowledges the hospitality of
Laboratorio Nacional de Fusi\'on, Asociaci\'on EURATOM-CIEMAT.


\appendix

\section{Riemann-Liouville fractional integral and differential operators}
\label{sec:FDOrealine}

The books \cite{OldSpa,Podlubny} are excellent introductory texts to
fractional calculus containing, in particular, the following
definitions.

Assume that $f:{\mathbb R}\to{\mathbb R}$ is a sufficiently
well-behaved function. The {\it Riemann-Liouville fractional integral
operators} of order $\alpha$ are defined as
\begin{eqnarray}\label{RL0}
&&{}_{a}D_x^{-\alpha} f:= \frac{1}{\Gamma(\alpha)}
\int_{a}^x (x-x')^{\alpha-1} f(x')\dd x',\cr
&&{}^{b}D_x^{-\alpha} f:=
\frac{1}{\Gamma(\alpha)}\int_{x}^{b}(x'-x)^{\alpha-1} f(x')\dd
x'.
\end{eqnarray}

As for the {\it Riemann-Liouville fractional differential operators}
of order $\alpha$, the definition is
\begin{eqnarray}\label{RL2}
&&{}_{a}D_x^\alpha f:=
\frac{1}{\Gamma(m-\alpha)}\frac{\dd^m}{\dd
x^m}\int_{a}^x\frac{f(x')}{(x-x')^{\alpha-m+1}}\dd x',\cr
&&~{}^{b}D_x^\alpha f:=
\frac{(-1)^{m}}{\Gamma(m-\alpha)}\frac{\dd^m}{\dd
x^m}\int_{x}^{b}\frac{f(x')}{(x'-x)^{\alpha-m+1}}\dd x',
\end{eqnarray}
where $m$ is the integer number verifying $m-1\le\alpha<m$. 

Finally, the Riesz fractional differential operator is defined as the
symmetric combination
\begin{equation}\label{eq:Riesz}
\frac{\partial^\alpha}{\partial |x|^\alpha} :=
\frac{-1}{2\cos(\pi\alpha/2)}\left({}_{-\infty}D^{\alpha}_x +
{}^{\infty}D^{\alpha}_{x}\right).
\end{equation}



\end{document}